\documentclass[pre,twocolumn,amsmath,amssymb,notitlepage]{revtex4-1}

\usepackage{dcolumn}
\usepackage{lipsum}
\usepackage{graphicx}
\usepackage[hidelinks]{hyperref}
\usepackage{float}
\usepackage{bm}
\usepackage[T1]{fontenc}
\usepackage[utf8]{inputenc}
\usepackage{color}
\usepackage{url}
\usepackage[bf]{subfigure}
\usepackage{rotating}
\usepackage[]{algorithm2e}
\usepackage{caption}

\SetAlgoCaptionSeparator{:}

\begin{document}

\title{Seeking Maximum Linearity of Transfer Functions}

\author{Filipi N. Silva$^1$}
\author{Cesar H. Comin$^1$}
\author{Luciano da F. Costa$^1$}
\affiliation{$^1$S\~{a}o Carlos Institute of Physics, University of S\~{a}o Paulo, PO Box 369, S\~{a}o Carlos, SP, Brazil}


\begin{abstract}
Linearity is an important and frequently sought property in electronics and instrumentation.   Here, we report a method capable of, given a transfer function, identifying the respective most linear region of operation with a fixed width.  This methodology, which is based on least squares regression and systematic consideration of all possible regions, has been illustrated with respect to both an analytical (sigmoid transfer function) and real-world (low-power, one-stage class A transistor amplifier) situations.  In the former case, the method was found to identity the theoretically optimal region of operation even in presence of noise.  In the latter case, it was possible to identify an amplifier circuit configuration providing a good compromise between linearity, amplification and output resistance.   The transistor amplifier application, which was addressed in terms of transfer functions derived from its experimentally obtained characteristic surface, also yielded contributions such as the estimation of local constants of the device, as opposed to typically considered average values. Moreover, the obtained characteristic surfaces of the considered transistor (a generic, low signal device) revealed a surprisingly complex structure.  The reported method and results paves the way to several other applications in other types of devices and systems, intelligent control operation, and other areas such as identifying regions of power law behavior.
\end{abstract}

\maketitle

\section{Introduction}

\begin{figure}[!htb]
  \begin{center}
  \includegraphics[width=0.95\linewidth]{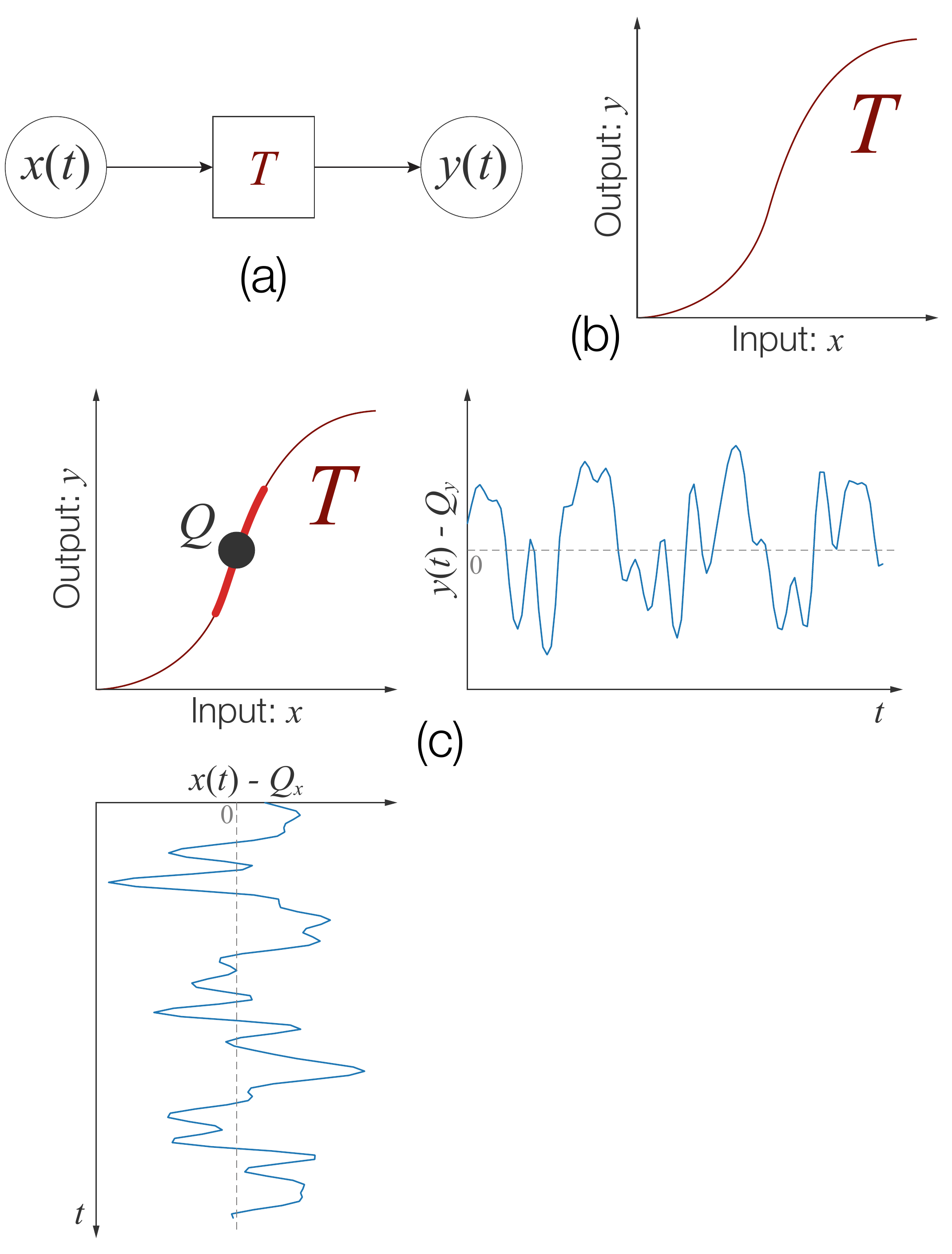}
  \caption{(a) Illustration of an input signal $x(t)$ being transformed by a system $T$ into an output signal $y(t)$. (b) The transfer function specifying the system. (c) It is often desired to have the operation point $Q=(Q_x,Q_y)$ of the system in the center of the most linear region of the transfer curve, so that the shifted input signal $x(t)-Q_x$ is transformed into the output signal $y(t)-Q_y$ with little distortion.}
  ~\label{f:transformFunction}
  \end{center}
\end{figure}

Several situations in applied sciences involve transforming a signal from an input to an output domain. This includes measuring any physical property through a sensor, conditioning a signal through a filter~\cite{Rabiner:1975aa} or amplifier~\cite{Rashid:2010aa}, and transducing an electrical signal into some action (e.g. a force). Any of these situations can be conveniently summarized in terms of a systems approach such as shown in Fig~\ref{f:transformFunction}(a), where the transforming system $T$ receives an input signal $x(t)$ and outputs a signal $y(t)$. The effect of the transformation can be clearly characterized in terms of the transfer function of the system, illustrated in Fig~\ref{f:transformFunction}(b).
 
Oftentimes, a linear mapping is desired between input and output, which ensures no modification, distortion or delay to the signal other than eventual scaling, value shifting or delay (linear phase).  Unfortunately, the linearity of real-world transfer functions are never perfect, being limited in several aspects, such as noise and distortions.  Yet, some of the regions of the transfer function are closer to being linear, and it becomes important to devise methods capable of selecting the best region for operation of the system.  Three main problems can be considered: (i) a maximum deviation from linearity $E_{max}$ is imposed on the sought region of a given length $L$ of the transfer function; (ii) given $L$ (along the input domain), find the region that minimizes the deviation from linearity; and (iii) given a maximum deviation from linearity $E_{max}$, the longest region is sought for the given transfer function.  In the former situation, the application requires a maximum acceptable distortion; in (ii), the objective is to select the best region of operation for a given application.   Observe that criterion (ii) is a particular case of (i), as it optimizes the error for the same required $L$.   In the present work, we concentrate on criterion (ii), which is often found in practice, in the sense that $L$ is pre-specified (e.g. in sensors and amplifiers applications, the desired output extension is a design imposition).  Such a methodology can be useful in best exploring the intrinsic capabilities of any sensor, amplifier or transducer, in the sense that maximum linearity operation can therefore be achieved for a given $L$, as illustrated in Fig.~\ref{f:transformFunction}(c).  Frequently, this region of interest is associated or defined by an operation (or quiescent) point $Q$, such as in Fig.~\ref{f:transformFunction}(c), which corresponds to the situation of the system under absence of signal (which defines the null level).  In most cases, the linear region should extend symmetrically along both sides from $Q$, in order to allow the maximum linearity.

Experimentally, the continuous transfer function of a system, sensor or transducer is never available, and needs to be sampled in terms of a sequence of points $S$.  The devised procedure (to be explained in detail in Section~\ref{s:method}) to find the most linear region for a given $S$ and $L$ performs minimization of the least mean square residues for several candidate regions.  The suggested procedure is evaluated in terms of the sigmoid function, which represents the transfer functions typically found in electronic systems~\cite{Lee:1996aa}. We note that an important assumption of the methodology is that the system is predominantly resistive.

To corroborate the practical usefulness of the introduced methodology, we apply it to the problem of determining the best operating points of a one-stage class A amplifier configuration~\cite{RCA:1973aa} based on a single generic low signal transistor.  We observe that we do not present a complete, operational, amplifier circuit, but only its analysis with respect to a resistive load.  The construction and analysis of high fidelity (hifi) audio systems, such as amplifiers, constitutes an important research subject nowadays because of the complexity of the concepts involved. In such systems, it is desirable to achieve the most linear transfer function from the original audio signal to the amplified output, leading to minimal distortions along the amplification process.  There are two main families of hifi audio amplifiers today, based on solid-state (e.g. transistors and integrated circuits) and vacuum tube technology, respectively.   Both these types of amplifiers can display distinct transfer curves depending on the parameters of the chosen components and the characteristics of the circuit, which are never underlain by fully linear relationships. Therefore, finding the best linear operation points paves the way to advancing in the designing and constructing hifi amplifiers.  The choice of a class A amplifier as a case example in this work is justified because this type of circuit is often appreciated by its linearity and simplicity, though typically at the expense of increased power consumption~\cite{self2013audio,cordell2011designing}.

The paper is organized as follows. Section~\ref{s:method} presents the in-detail description of the proposed methodology to obtain the most linear regions of a transfer curve. In section~\ref{s:sigmoid} the methodology is illustrated and validated with respect to a set of Sigmoid functions. In section~\ref{s:amplifier} we illustrate the application of the methodology to a real device (a generic small signal BJT).  We start from the experimentally obtained characteristic surface and then quantify the performance of several possible derived load lines.

\section{Methodology}
\label{s:method}
\begin{figure}[!htb]
  \begin{center}
  \includegraphics[width=1.0\linewidth]{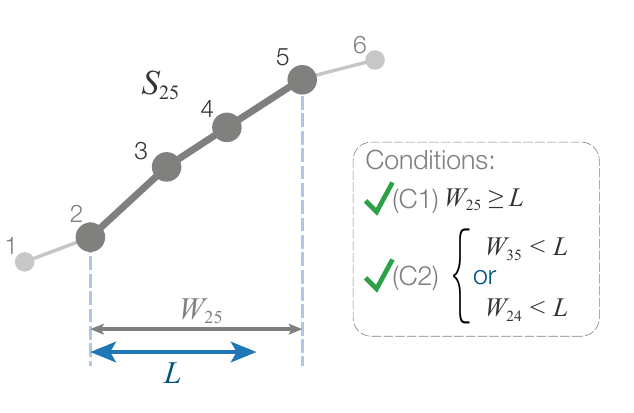} 
  \caption{Example of subsequence having a valid target size $L$. Given the original sequence of six points, the highlighted subsequence $S_{2,5}$ has size $W_{2,5}\geq L$, which passes condition (C1). When removing point 2 or 5, the size of the subsequence becomes smaller than $L$, which is in agreement with condition (C2).}
  ~\label{f:conditions}
  \end{center}
\end{figure}

In the following, we consider a given sequence of points $S=((x_1,y_1),(x_2,y_2),\dots,(x_n,y_n))$, describing the relationship between variables $x$ and $y$. An example of such a sequence is shown in Figure~\ref{f:conditions}. Although we consider $S$ to be a generic sequence, it can have different meanings, such as data sampled from a known continuous function or from an experiment. A contiguous subsequence $S_{k,q}$ of $S$ is defined as the sequence of $m=q-k+1$ points in $S$ having index $i$ in the range $[k,q]$~\footnote{Specifically, the subsequence is given by $S_{k,q}=((x_k,y_k),(x_{k+1},y_{k+1}),\dots,(x_{q=k+m-1},y_{q=k+m-1}))$}. 

As mentioned in the previous section, the linearity of the transfer curve of a system (e.g. sensor, filter, amplifier) should be optimal in the expected operation range $L$ of the system. Therefore, we only consider subsequences $S_{k,q}$ having a size $W_{k,q}=x_q-x_k$ that is as close as possible to the desired target range $L$. This is done by selecting subsequences contained in $S$ that obey the following conditions:

\begin{align*}
\mathbf{C1:} & \,\,\,\, W_{k,q}\geq L \\
\mathbf{C2:} & \begin{cases} 
W_{k+1,q}<L \\ 
\mbox{or} \\
W_{k,q-1}<L
\end{cases}
\end{align*}

These conditions are illustrated in Figure~\ref{f:conditions}. The subsequence $S_{2,5}$ shown in the figure follows both conditions because its size is larger than $L$ (condition 1) and, after removing one of its endpoints, its size becomes smaller than $L$ (condition 2). Subsequences that follow these two criteria are considered valid for linearity quantification. 

\begin{figure*}[!htb]
  \begin{center}
  \includegraphics[width=0.95\linewidth]{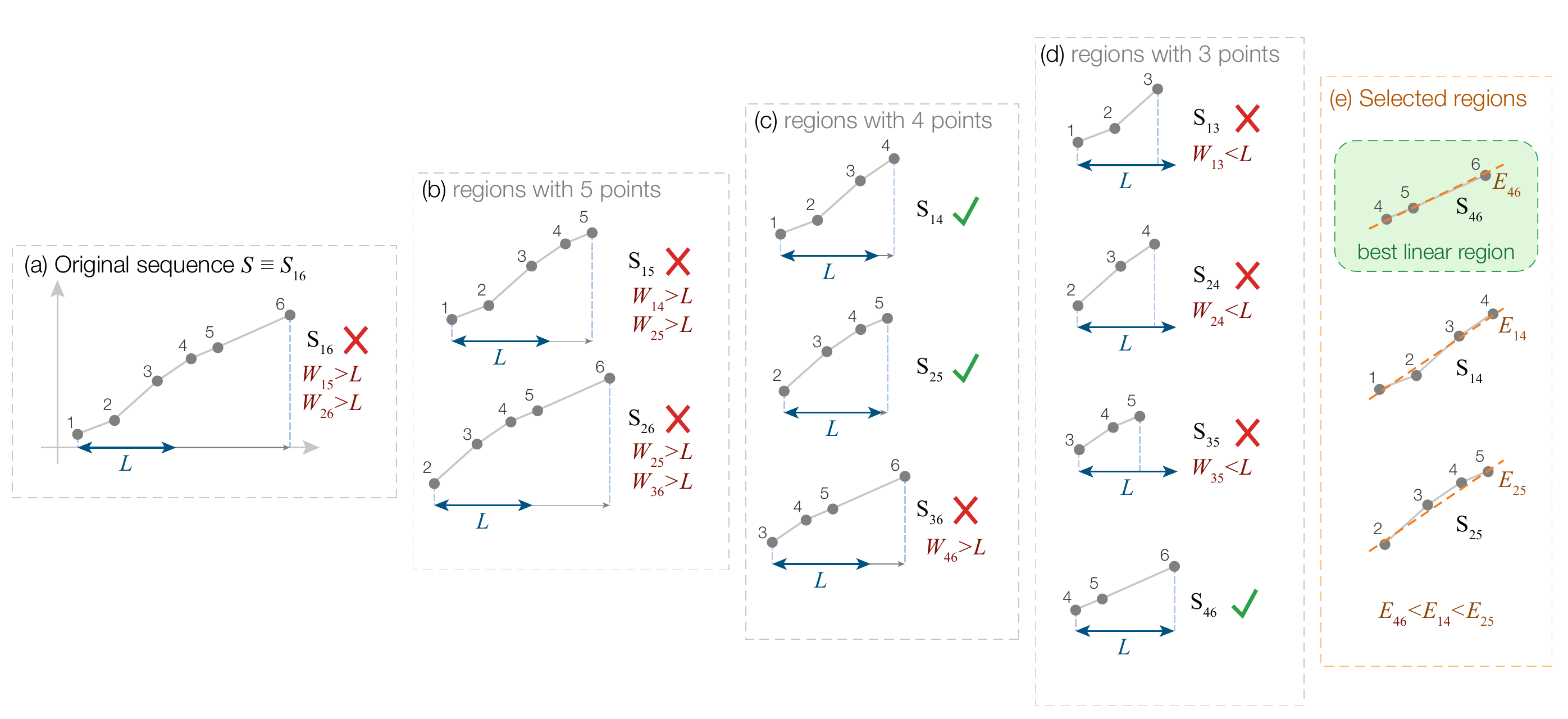} 
  \caption{Example of application of the methodology. The original sequence $S$, containing six points, is shown in (a). All subsequences of $S$ with at least $3$ points are considered for the initial selection. Sequences with 5, 4 and 3 points are presented in, respectively, (b), (c) and (d). The target range $L$ is indicated below each subsequence. Check marks indicate subsequences that comply with conditions (C1) and (C2), while discarded subsequences are marked with an {\bf X}. The selected subsequences are shown in (e), where the most linear subsequence, i.e., the one having the lowest residue $E_{k,q}$, is highlighted.}
  ~\label{f:method}
  \end{center}
\end{figure*}

In order to asses how linear a given subsequence is, we need to quantify the deviation, $E$, of such subsequence from a straight line. This deviation can have different definitions. One traditional approach is to calculate the sum of the squared distances, in the $y$ coordinate, between the points and a candidate straight line adjusted to the data~\cite{bevington2003data}. The process of finding the straight line that minimizes the sum of squared distances is known as linear least squares regression~\cite{draper1966applied,bevington2003data}, and the respective error of the linear regression can be used to quantify the linearity of the points in a candidate subsequence. This error is given by

\begin{equation}
E_{k,q}=\sqrt{\frac{1}{m}\sum\limits_{i=k}^{q}\left(y_i-\alpha x_i-\beta\right)^2},
\end{equation}
where $\alpha$ and $\beta$ are, respectively, the slope and the $y$ intercept value of the best-fitting linear function. 

The methodology consists in applying the linear least squares regression to all subsequences of $S$ following conditions (C1) and (C2). A simple, but not optimal, approach for such a task is to explore all existing subsequences in the investigated sequence of points $S$. This can be done by varying both $k$ and $q$, such that $1\leq k<q\leq |S|$, and checking if the resulting range $[k,q]$ follows the aforementioned conditions. In addition, all subsequences must have at least $3$ data points for the analysis, since $2$ points always define a linear subsequence. We note that the process can be optimized by preemptively discarding ranges containing subsequences that were already considered valid for linearity quantification. A linear least squares regression is then applied to the points belonging to each valid subsequence $S_{k,q}$. Next, the respective error, $E_{k,q}$, of each regression is calculated. Finally, the subsequence associated with the lowest error defines the most linear extent of $S$. Figure~\ref{f:method} illustrates the application of the methodology to a small sequence of points. In the figure, all possible subsequences (ten in total) that can be applied to the sequence of six points are shown. A check mark is used to indicate subsequences that follow the aforementioned conditions.

Algorithm~\ref{a:bestlinearalg} summarizes the process of finding the best linear region in a sequence of points $S$ for a given $L$. The function {\bf bestLinearFitError}($S_{k,q}$) calculates the residue obtained when applying the least squares method to the subsequence $S_{k,q}$.

\begin{algorithm}[!h]
\SetKwData{Left}{left}\SetKwData{This}{this}\SetKwData{Up}{up}
\SetKwFunction{Union}{Union}\SetKwFunction{FindCompress}{FindCompress}
\SetKwInOut{Input}{input}\SetKwInOut{Output}{output}
\Input{A sequence of points $S=((x_1,y_1),(x_2,y_2),\dots,(x_n,y_n))$.}
\Input{Minimum region length $L$.}
\Output{A tuple $[k_\mathrm{best},q_\mathrm{best}]$ corresponding to the best fitted subsequence and its residue $E_\mathrm{best}$}
\BlankLine
$\text{windows} \leftarrow [\,]$\;
$E_\mathrm{best} \leftarrow \infty$\;
$k_\mathrm{best} \leftarrow \varnothing$\;
$q_\mathrm{best} \leftarrow \varnothing$\;
\For{$k \leftarrow 1$ \KwTo $|S|$}{
  \For{$q \leftarrow (k+2)$ \KwTo $|S|$}{
     \If{$[k,q]$ {\bf contains no ranges of} windows}{
       \If{$x_{q}-x_{k}  \geq L$ \\ {\bf and} $(x_{q-1}-x_{k} < L$ {\bf or} $x_{q}-x_{k+1} < L)$}{
          {\bf append} $[k,q]$ {\bf to} windows\;
          $E \leftarrow$ {\bf bestLinearFitError}($S_{k,q}$)\;
          \If{$E < E_\mathrm{best}$}{
               $E_\mathrm{best} = E$\;
             $k_\mathrm{best} = k$\;
             $q_\mathrm{best} = q$\;
             }
          }
       }
     }
   }
   \caption{\label{a:bestlinearalg} Algorithm to determine the best linear region of a sequence $S$ for given length $L$.}
\end{algorithm}


\section{Linearity on artificial data}
\label{s:sigmoid}

\begin{figure}[]
  \begin{center}
  \includegraphics[width=0.95\linewidth]{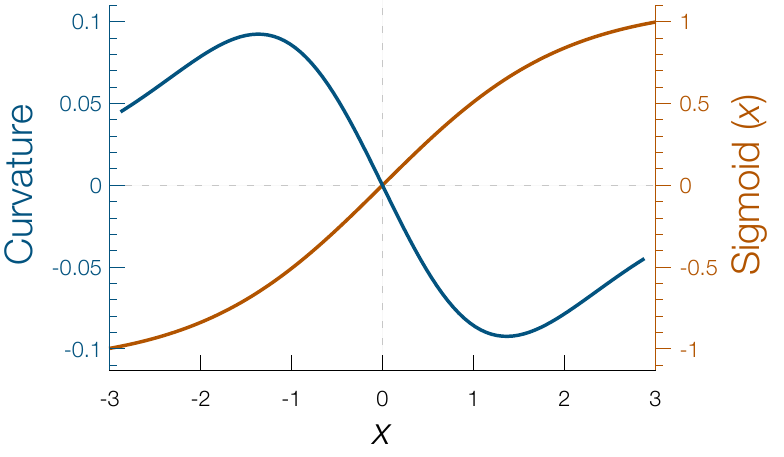} 
  \caption{The logistic function and its respective local curvature.}
  ~\label{f:SigmoidCurvature}
  \end{center}
\end{figure}

In order to illustrate the potential of the methodology to quantify linearity, in this section we present the application of the methodology to a sigmoid function. For such a task, we considered the logistic function, given by

\begin{equation}
f(x)=\frac{1}{1+e^{-x}}.
\end{equation}
This function was chosen because it has a clear linear region around $x=0$, while the non-linearity of the function increases with $|x|$, until reaching saturation. This behavior is indicated in Figure~\ref{f:SigmoidCurvature}, where we plot the logistic function and its respective curvature~\cite{da2009shape}. Note that we considered the interval $[-3,3]$ for the function domain. The plot shows that at $x=0$ the curvature is zero, meaning that the function is locally linear at this point. The curvature increases when going away from $x=0$, until it starts to decrease again since the logistic function tends to a constant value for $|x|\rightarrow \infty$, due to saturation. Therefore, $x=0$ should represent the optimal operation point of a logistic transfer function as far as linearity is concerned.

In order to verify the robustness of the methodology for identifying linear regions, we added different levels of noise to the function $f(x)$. Since $f(x)$ has its values defined in the interval $[0,1]$, the noise level is represented as a fraction $r$ of this interval, or equivalently, as a percentage $100r$ of the function range. Given a noise level $r$, we define a new function 

\begin{equation}
g_r(x)=f(x)+\zeta(x),
\end{equation}
where $\zeta$ is a random variable having a uniform distribution in the interval $[-r/2,r/2]$. In such a case, the region near the origin should be considered the most linear by the methodology.

\begin{figure*}[]
  \begin{center}
  \includegraphics[width=0.70\linewidth]{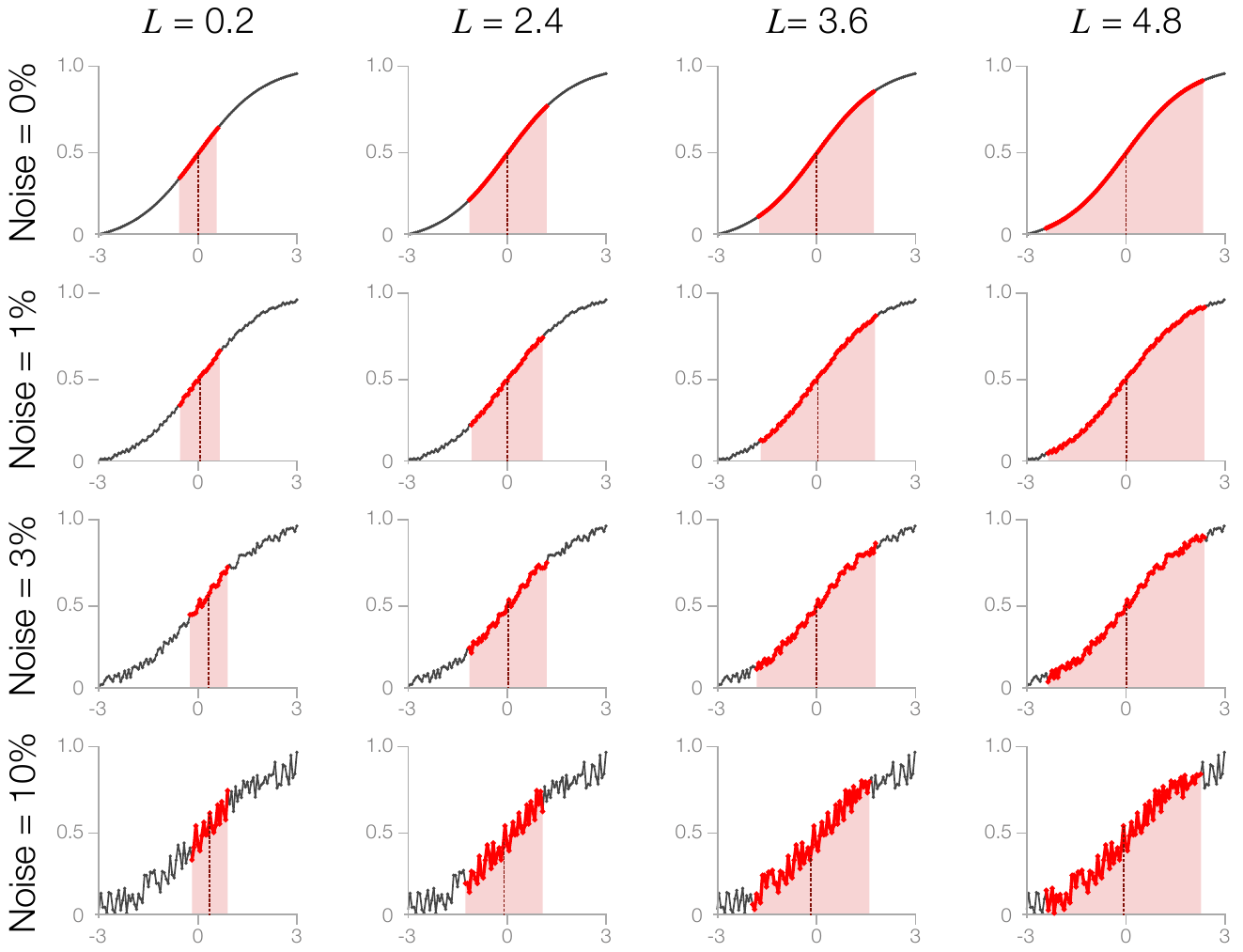} 
  \caption{Identification of the most linear region of the logistic function. Each row of plots is respective to the logistic function having different noise levels, while each column contains the results for a distinct minimum window size $L$. Regions marked in red represent the most linear interval found by the method.}
  ~\label{f:Sigmoids}
  \end{center}
\end{figure*}

We tested the methodology for different noise levels $r$ and distinct values for the minimum range $L$. The results are shown in Figure~\ref{f:Sigmoids}. Each row of plots correspond to a distinct noise level, while each column is respective to a different $L$. The largest linear region of each considered case is indicated in red. The results show that the methodology identifies the region near the origin as being the most linear, as expected by the properties of the logistic function, as well as by a visual inspection of the function shape. We observe a small variation on the central position of the most linear region when $L$ is comparable to the noise level added to the function. Therefore, the results indicate that the methodology is robust against random perturbations on the analyzed function.

\begin{figure*}[]
  \begin{center}
  \includegraphics[width=0.60\linewidth]{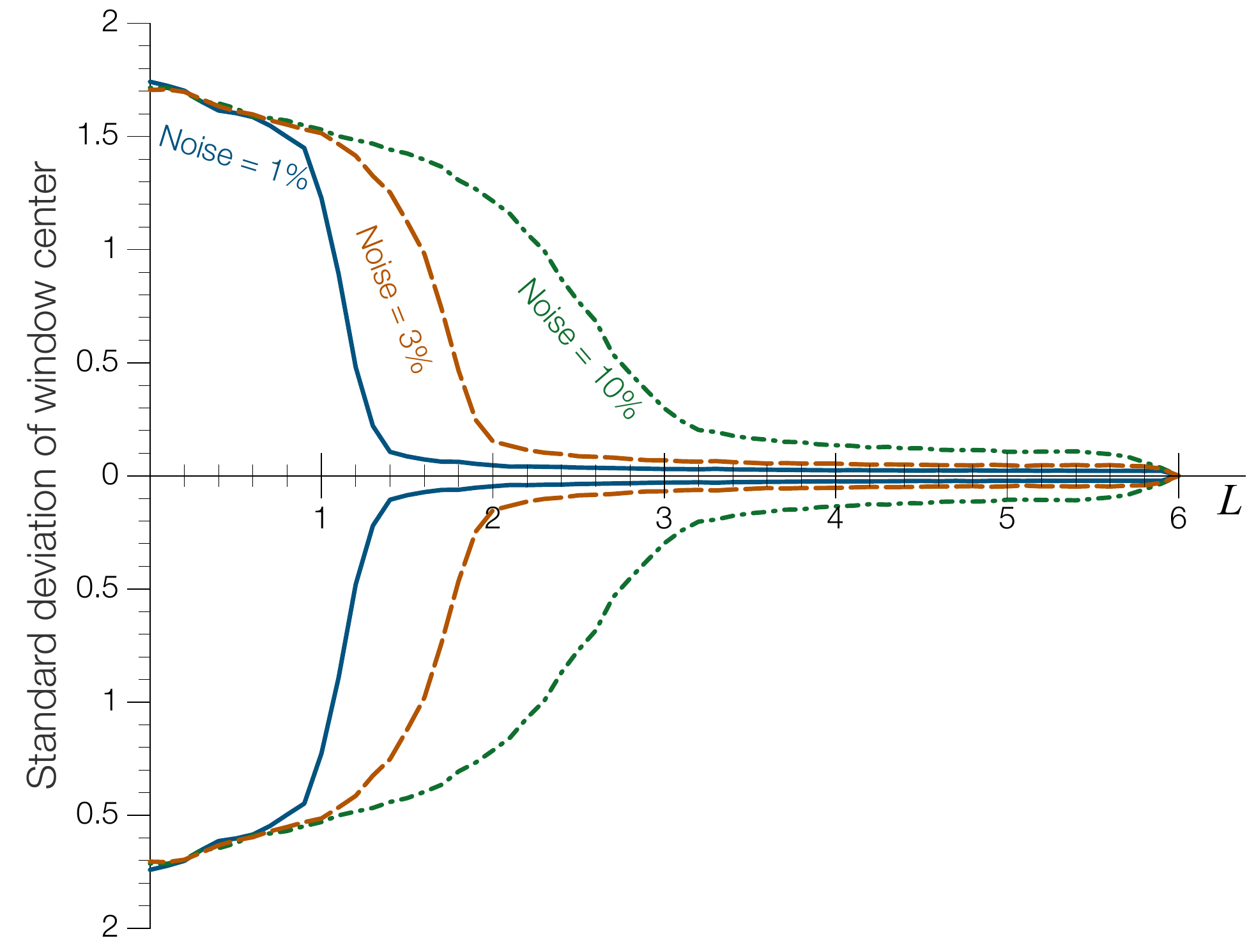} 
  \caption{Standard deviation of the most linear region of the noisy logistic function, as a function of the minimum window size $L$. Each line represents a different noise level $r$ added to the data.}
  ~\label{f:SigmoidCenters}
  \end{center}
\end{figure*}

In order to generalize the results obtained when applying the methodology to the logistic function, we considered distinct realizations of the noise $\zeta$ added to function $f(x)$, and calculated the optimal operation point for each realization. Then, the respective standard deviation of the calculated positions was estimated, for different values of $L$. The results are shown in Figure~\ref{f:SigmoidCenters}. Each curve in the plot is relative to a distinct noise level $r$, as indicated. The plot shows that the position of the most linear region can have large changes depending on the noise level and the parameter $L$. Still, the position always tend to 0 for large $L$, showing that a proper choice of the minimum range is important for the methodology.

\section{Case example: Class A one-stage transistor amplifier}
\label{s:amplifier}

Given their ability to change the amplitude of electronic signals, amplifiers are involved in many electronic systems.  In particular, audio amplifiers play a critical role in transforming the low power audio signals generated by the source (e.g. CD player, DAD, etc.) into audible sound.  In a high fidelity (hifi) system, the amplifier should only uniformly affect the amplitude of the input signal, which requires a nearly linear transfer function covering the respective operation region.  Typically, several stages are required in order to accomplish the desired amplification, which demands special care in achieving good linearity levels at each stage.   Here, we consider analog audio amplifiers, particularly those in the class A, which is characterized by 100\% of the signal being used~\cite{self2013audio,shea1955transistor}.

\begin{figure}[]
  \begin{center}
  \includegraphics[width=1.0\linewidth]{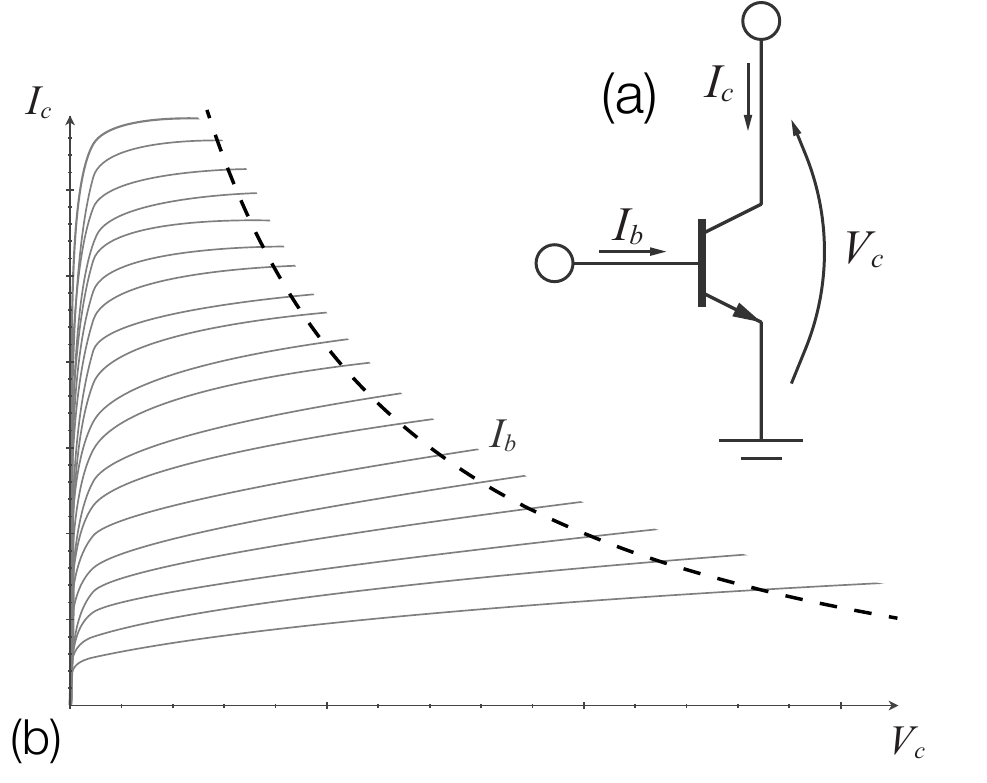} 
  \caption{A generic NPN BJT (a) and the characterization of its properties in terms of isolines in the $V_c\times I_c$ space (b).
                The maximum dissipation power is shown by the dashed curve.}
  ~\label{f:curvesValve}
  \end{center}
\end{figure}

In this work we tackle the study of a one-stage class A amplifier using a low signal BJT (bipolar junction transistor).  The schematics of an NPN BJT is shown in Figure~\ref{f:curvesValve}(a).  Typically, transistor amplifiers incorporate high degree of feedback, which reduces the effect of wide variability of real-world transistor constants such as $\beta$~\cite{shea1955transistor,palumbo2007feedback}.  However, in the present work we consider a less common circuit devoided of feedback, so as to provide a more diversified operation and linearity as the circuit parameters are varied, therefore allowing a better validation of the proposed linearity method.

In this section, we apply the method proposed in Section~\ref{s:method} to the problem of choosing the operation point of a one stage class A amplifier in order to maximize linearity, given a  desired input range.  For generality's sake, we are not restricted to finding the best configuration along a load line, instead we consider many putative load lines derived from the characteristic surface defining the device operation.  In other words, given the device characteristics, the range of operation, and type of circuit, the reported methodology is capable of identifying the best operation point.  First, experimental data is obtained and interpolated as the characteristic surface, in order to allow accurate estimation of the partial derivatives required for modeling the transfer function.  We show that different operation points lead to varying compromises between output resistance, current gain and linearity.

Mathematically, the transistor operation can be described in terms of the state variables $I_c(I_b,V_c)$, $I_b(I_c,V_c)$ and $V_c(I_b,I_c)$. Where $I_b$ and $I_c$ are, respectively, the input and output currents of the transistor and $V_c$ the collector voltage~\cite{zimmerman1959electronic}. Therefore, a given transistor has a well-defined surface in the $I_c\times I_b\times V_c$ space, defined by the relationship between these three properties. It is common practice to visualize such a surface as isolines in a 2D $V_c\times I_c$ space. An example of such visualization is shown in Figure~\ref{f:curvesValve}(b).

The $S(I_c, I_b, V_c)$ surface properties of a transistor are specified by a set of so-called constants, referred as \emph{current gain} ($\beta$), \emph{transresistance} ($R_T$) and \emph{output resistance} ($R_o$). These constants can be defined in terms of the partial derivatives of the transistor state variables, that is

\begin{align}
\beta & = \frac{\partial I_c}{\partial I_b} \label{eq:beta}\\
R_T & = \frac{\partial V_c}{\partial I_b}\label{eq:rt}\\
R_o & = \frac{\partial V_c}{\partial I_c}\label{eq:ro}.
\end{align}

The current gain is associated with the expected transistor current amplification, but the actual amplification of the circuit also depends on the parameters of the latter. The transresistance indicates the differential voltage variation of the transistor output for a small variation of $I_b$. Therefore, a small value of this property is useful to minimize undesired effect from the reactive components in the circuit. The output resistance influences the transfer of power to the load.

\begin{figure*}[]
  \begin{center}
  \includegraphics[width=0.85\linewidth]{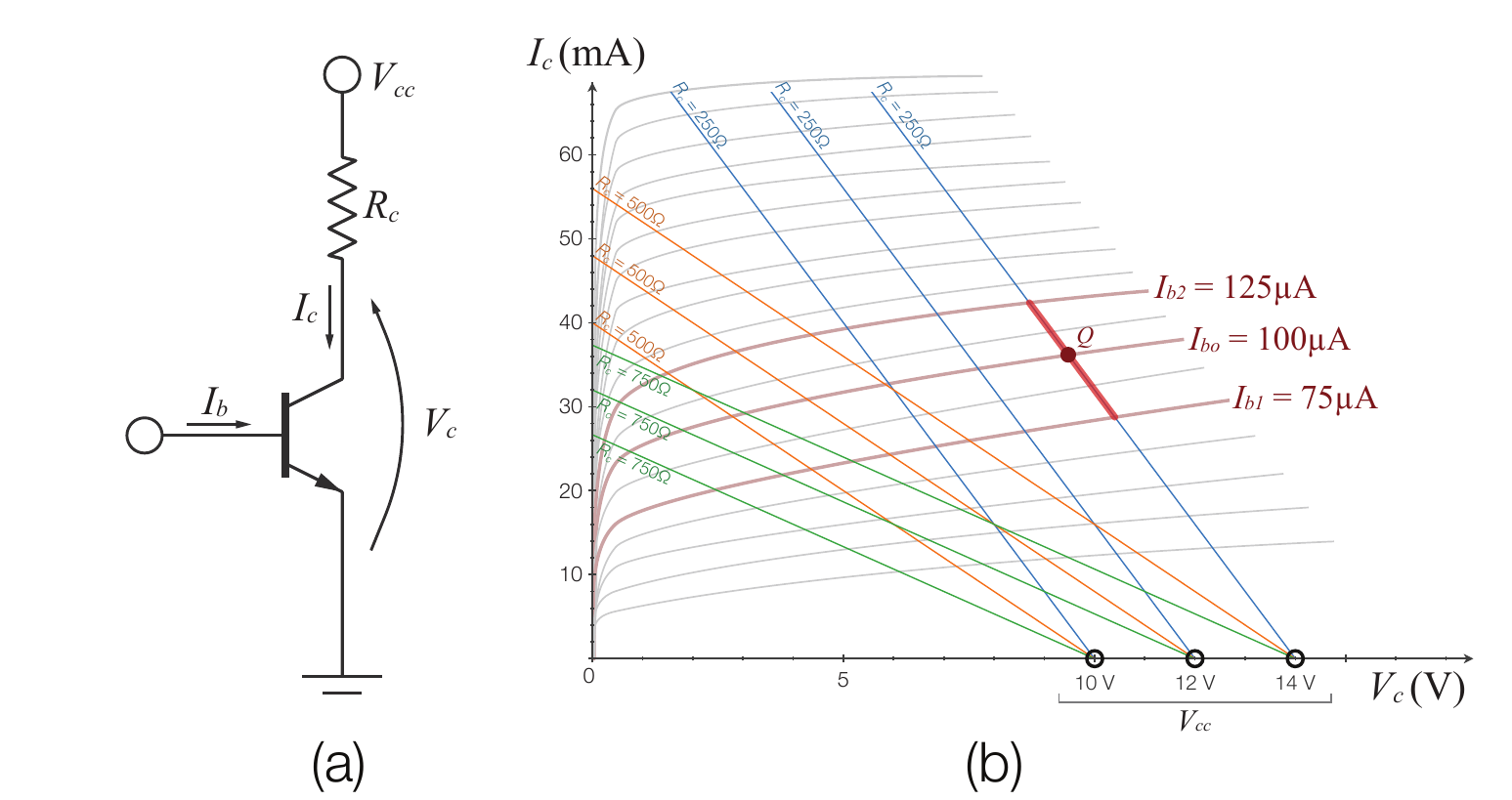} 
  \caption{(a) Circuit of a one-stage Class A amplifier with resistive load (a).  Load lines defined by distinct values of $V_{cc}$ and $R_c$, shown in the $V_c\times I_c$ plane (b).}
  ~\label{f:tubeCircuit}
  \end{center}
\end{figure*}

The basic circuit for the one-stage class A amplifier is shown in Figure~\ref{f:tubeCircuit}(a).  In order to simplify the analysis, we consider a purely resistive load.  This circuit has two parameters, the main power supply ($V_{cc}$) and the resistance ($R_c$). These two parameters define a load line for the transistor, which restricts the relationship between $I_c$ and $I_b$ to a line in the $V_c\times I_c$ plane.  Examples of load lines are shown in Figure~\ref{f:tubeCircuit}(b).   Also shown in Figure~\ref{f:tubeCircuit}(b), in particular for the rightmost load line, is a specific configuration of operation point defined by $I_{bo} = 100\mu A$ as well as a region of operation extending between $I_{b1} = 75 \mu A$ to $I_{b2} = 125 \mu A$.  Observe that the operation point is defined by the intersection between the load line and the isoline $I_{bo}$.  A relevant property of the circuit is the total current gain, $A$, for a given region of operation defined as 
\begin{equation}
A = \frac{dI_c}{dI_b}.
\end{equation}
This property describes the actual current amplification imposed by the circuit for given BJT constants and circuit parameters. Please refer to Appendix A for more information.   Typically, the aim of a hifi amplifier is to provide a linear relationship between $I_b$ and $I_c$ for a selected load line.

\begin{figure*}[]
  \begin{center}
  \includegraphics[width=0.75\linewidth]{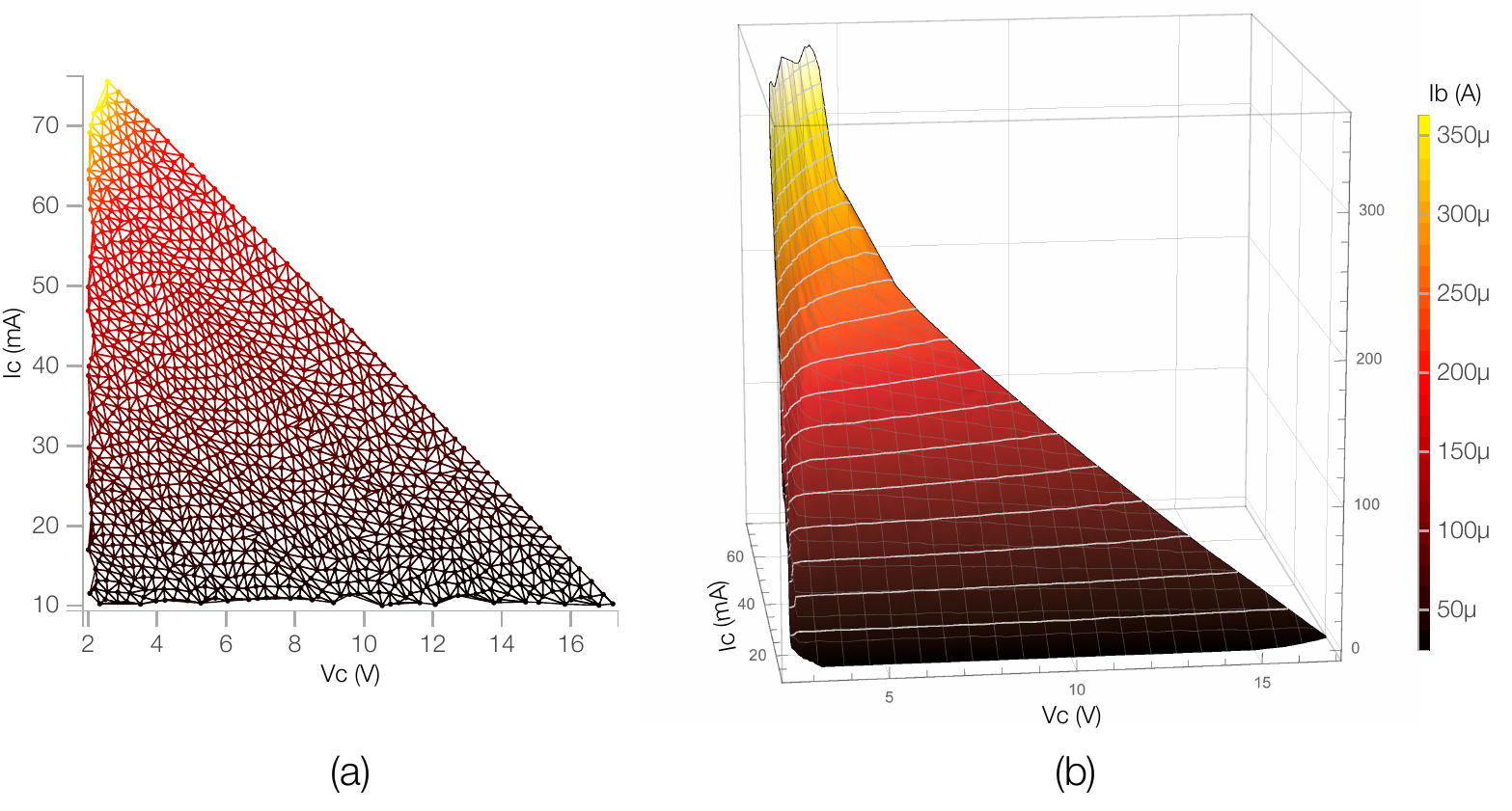} 
  \caption{Interpolated relationship between circuit properties $I_c$, $V_c$ and $I_b$. (a) Delaunay triangulation of the experimental data and (b) the respective interpolated surface in the three-dimensional space.}
  ~\label{f:IaVaVgSurface}
  \end{center}
\end{figure*}

In order to obtain the $S(V_c, I_c, I_b)$ surface, we experimentally sampled the $I_c (V_c) $ curves along load lines with fixed $R_c$ for a sequence of $V_{cc}$ values. These isolines were then used to interpolate $I_b$ along the plane $V_c\times I_c$. This was accomplished by first obtaining a uniform sampling of the experimental points. Next, the Delaunay triangulation technique~\cite{preparata1985shamos} was used to derive a mesh of triangles lying in the $I_c\times V_c\times I_b$ space (shown in Figure~\ref{f:IaVaVgSurface}(a)). The continuous function used to define the $S(V_c, I_c, I_b)$ surface was then obtained by applying the baryocentric interpolation method~\cite{Skala2008120} over the mesh of triangles. The obtained surface is shown in Figure~\ref{f:IaVaVgSurface}(b).

\begin{figure*}[]
  \begin{center}
  \includegraphics[width=0.95\linewidth]{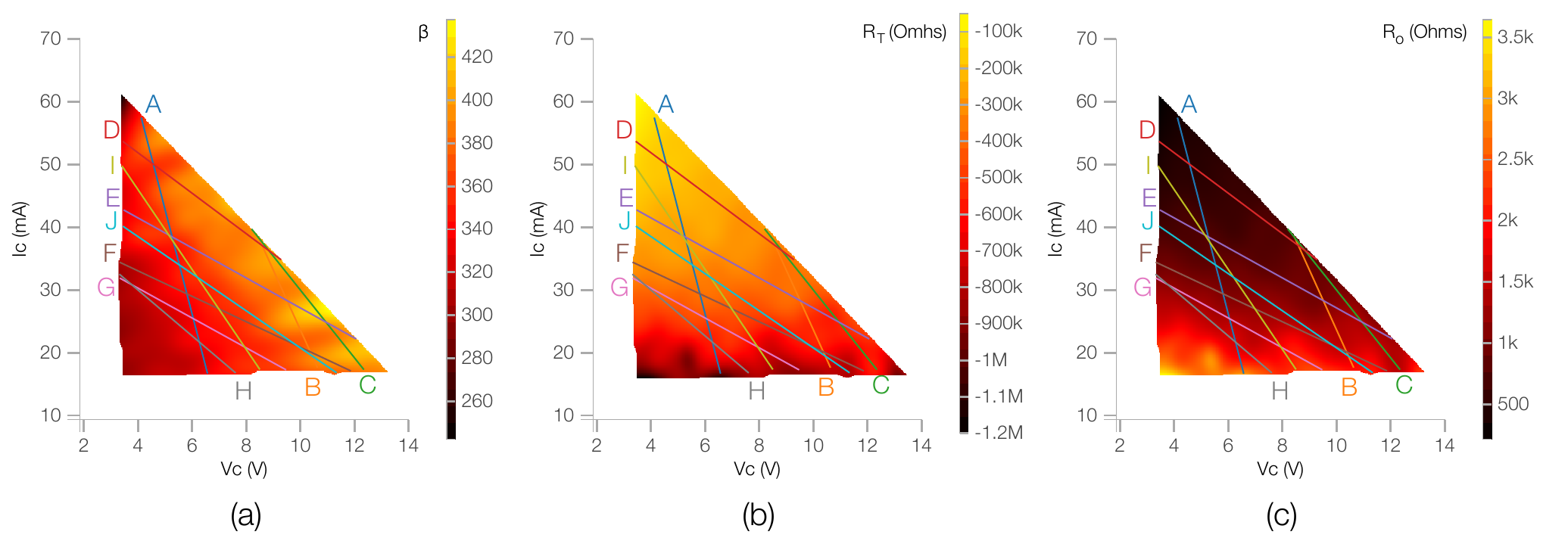} 
  \caption{Transistor constants calculated from the interpolated values shown in Figure~\ref{f:IaVaVgSurface}. The constants, and the respective equations defining them, are (a) current gain (Equation~\ref{eq:beta}), (b) transresistance (Equation~\ref{eq:rt}) and (c) output resistance (Equation~\ref{eq:ro}). Some load lines considered throughout the discussion are shown in each figure.} 
  ~\label{f:IaVaMaps}
  \end{center}
\end{figure*}

The obtained surface is smooth enough to allow differentiation. Therefore, from this surface we estimated three relevant constants of the transistor, i.e. $\beta$, $R_o$ and $R_T$. The results, shown in Figure~\ref{f:IaVaMaps}, provide a much more informative characterization of these three properties than the minimum and maximum values typically given in transistor datasheets.  The region shown in this figure, which corresponds to the circuit configurations covered by the experimental results and interpolation, is henceforth called \emph{polyhedron}. The $\beta$ values vary from $242$ to $437$, with average 360 and standard deviation 30,  reaching its highest values at the right lower region of the polyhedron in Figure~\ref{f:IaVaMaps}(a). The obtained $R_T$ values, depicted in Figure~\ref{f:IaVaMaps}(b), range from $-1.2M\Omega$ to $-50k\Omega$, peaking at the upper corner of the polyhedron. An opposite trend is verified for $R_o$, as shown in Figure~\ref{f:IaVaMaps}(c). The surfaces obtained for these three transistor constants present some lump-like irregularities, which are in agreement with the variation of beta suggested by the changing slopes of experimental isolines sometimes found in the literature (e.g.~\cite{shea1955transistor}).

\begin{figure*}[]
  \begin{center}
  \includegraphics[width=0.80\linewidth]{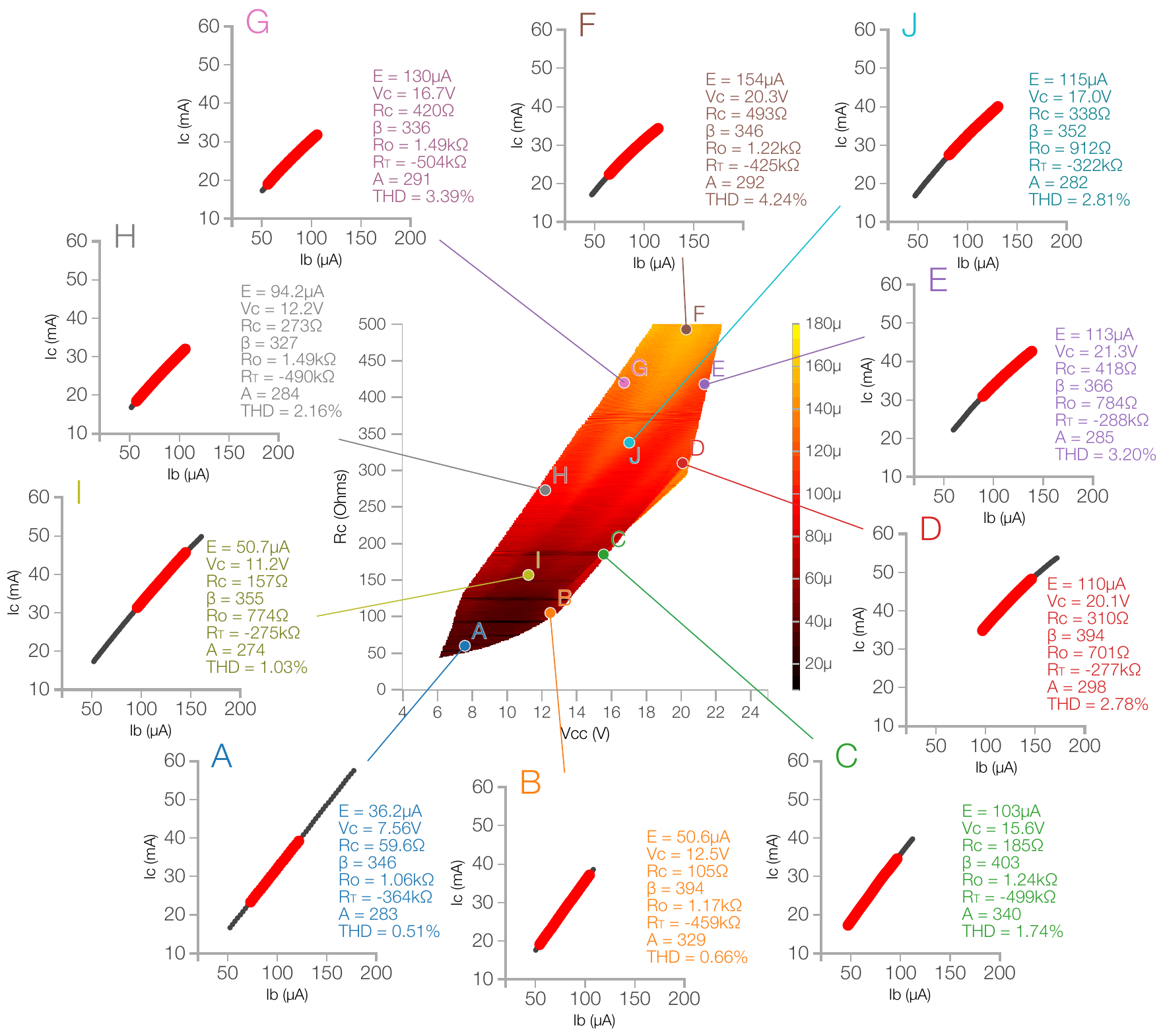} 
  \caption{Heat map of the linearity error for distinct values of $V_{cc}$ and $R_c$. The relationship between the input current, $I_b$, and output current, $I_c$, is shown for the chosen load lines.}
  ~\label{f:ErrorLinearity}
  \end{center}
\end{figure*}

As mentioned above, each pair of circuit parameters $(V_{cc}, R_c)$ implies a load line that defines the operation of the circuit. Some examples of load lines are shown in Figure~\ref{f:IaVaMaps}. The systematic variation of parameters $V_{cc}$ and $R_c$ allows a throughout analysis of the circuit properties at distinct operation conditions. These parameters are bounded by the adopted values of the transistor constants shown in Figure~\ref{f:IaVaMaps}. By considering all these allowed values of $V_{cc}$ and $R_c$ we can define an operation domain $\mathcal{S}$ for the circuit.  The considered load lines are specified by sampling this domain with 500 points of resolution for each of the circuit parameters.  The methodology presented in Section~\ref{s:method} was applied to each considered load line, given a target input range of $L=50\mu\text{A}$. The resulting linearity error, $E$, over $\mathcal{S}$ is shown in Figure~\ref{f:ErrorLinearity}.  It is clear that the error increases steadily upwards along the vertical.  The most linear regions are to be found precisely for low values of $R_c$ and $V_{cc}$.  In Figure~\ref{f:ErrorLinearity} we also show the transfer curves defined by a few chosen load lines.  These load lines were chosen as they were found to provide a good representation of the circuit properties inside domain $\mathcal{S}$, since the linearity shows smooth variation along $\mathcal{S}$.   The selected operation range $L$ of each transfer curve is indicated in red. We note that the selected load lines are the same as those indicated in Figure~\ref{f:IaVaMaps}. The four first columns of Table~\ref{t:parVals} present the values of $V_{cc}$, $R_c$ and $E$ for each of these load lines, specified by labels.

\begin{table}[]
\centering
\footnotesize
\caption{Throughout the discussion we consider some particularly interesting load lines for the circuit. The table shows the relevant properties of such lines. These properties, and their respective physical units, correspond to the circuit voltage ($V_{cc}$, in Volts), circuit resistance ($R_c$, in $\Omega$), linearity error ($E$, in $\mu A$), average circuit amplification ($A$), average output resistance ($R_o$, in $\text{k}\Omega$), average current gain ($\beta$), average transresistance ($R_T$, in $\Omega$), and total harmonic distortion (THD, in percentage).}
\label{t:parVals}
{\setlength{\tabcolsep}{0.18cm}
\begin{tabular}{ c c c c c c c c c}
\hline
Label  & $V_{cc}$ & $R_c$ & $E$ & $\langle A\rangle$  & $\langle R_o\rangle$ & $\langle\beta\rangle$ & $\langle R_T\rangle$ & THD \\ \hline
A & 7.56 & 59.6 & 36.2 & 283  & 1.06      & 346 & -364  & 0.51\% \\
B & 12.5 & 105 & 50.6 & 329   & 1.17      & 394 & -459   & 0.66\% \\
C & 15.6 & 185 & 103 & 340    & 1.24      & 403 & -499    & 1.74\% \\
D & 20.1 & 310 & 110 & 298    & 0.701    & 394 & -277      & 2.78\% \\
E & 21.3 & 418 & 113 & 285    & 0.784    & 366 & -288       & 3.20\% \\
F & 20.3 & 493 & 154 & 292     & 1.22      & 346 & -425     & 4.24\% \\
G & 16.7 & 420 & 130 & 291    & 1.49      & 336 & -504   & 3.39\% \\
H & 12.2 & 273 & 94.2 & 284   & 1.49      & 327 & -490   & 2.16\% \\
I & 11.2 & 157 & 50.7 & 274     & 0.774    & 355 & -275     & 1.03\% \\
J & 17.0 & 338 & 115 & 282     & 0.912    & 352 & -322     & 2.81\% \\ \hline
\end{tabular}}
\end{table}

\begin{figure*}[]
  \begin{center}
  \includegraphics[width=0.70\linewidth]{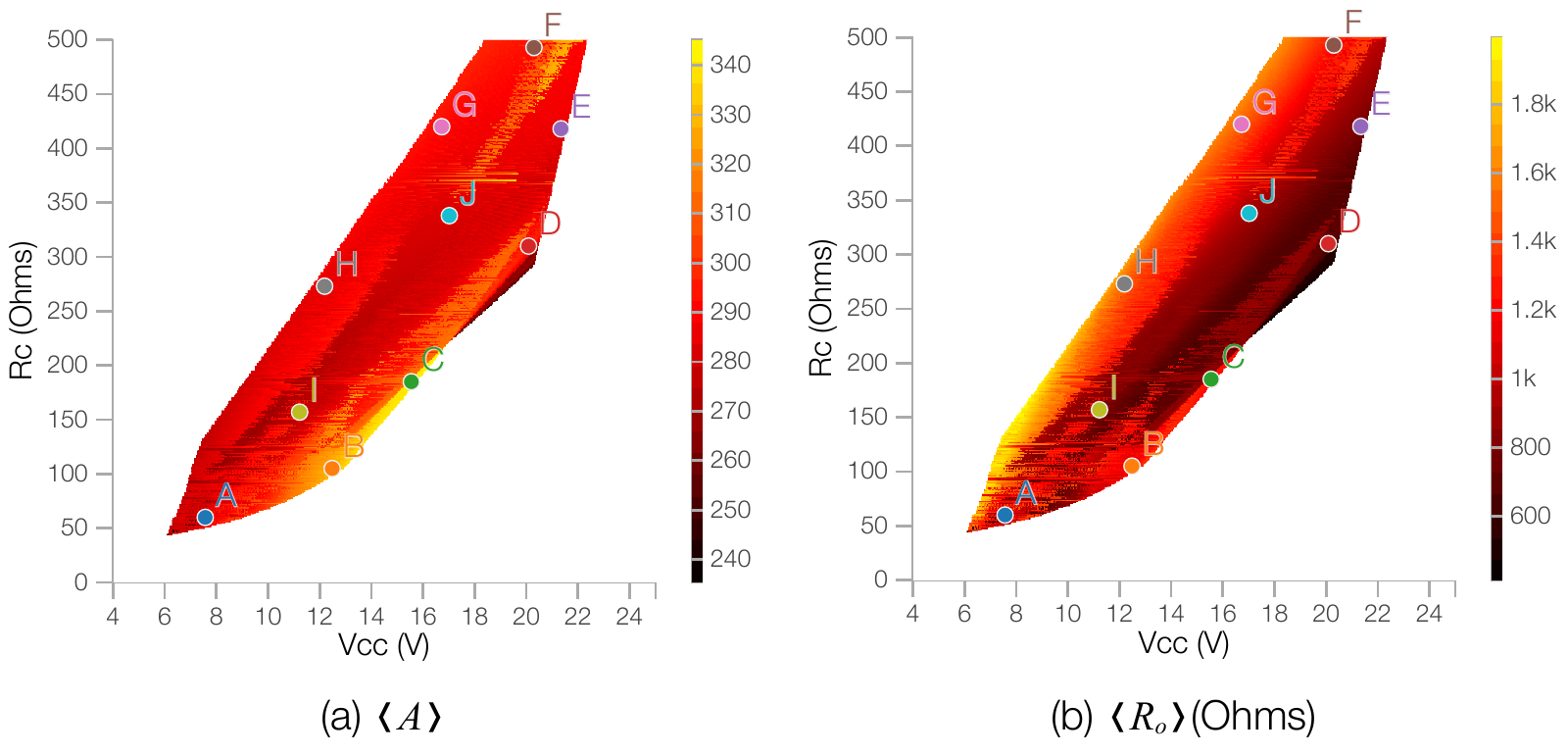} 
  \caption{Circuit and transistor properties calculated for distinct values of $V_{cc}$ and $R_c$. (a) Average circuit amplification and (b) output resistance. We also show in each plot the position of the load lines indicated in Figure~\ref{f:IaVaMaps}.}
  ~\label{f:RlHTMap}
  \end{center}
\end{figure*}

Besides the requirement that a proper load line should provide a highly linear relationship between $I_c$ and $I_b$, other properties of the circuit are often also sought. For instance, one may seek a large amplification and/or low output resistance, the latter being typically useful to minimize the influence of reactive loads.  In Figures~\ref{f:RlHTMap}(a) and (b) we show the averages of, respectively, the amplification, $\langle A \rangle$, and output resistance, $\langle R_o \rangle$, obtained for the chosen load lines. The averages were calculated along the respective operation range found by the linearity methodology for each load line. The labels identify the chosen load lines.  The values of the average amplification, output resistance, current gain and transresistance are indicated in Table~\ref{t:parVals}.

\begin{figure*}[]
  \begin{center}
  \includegraphics[width=0.75\linewidth]{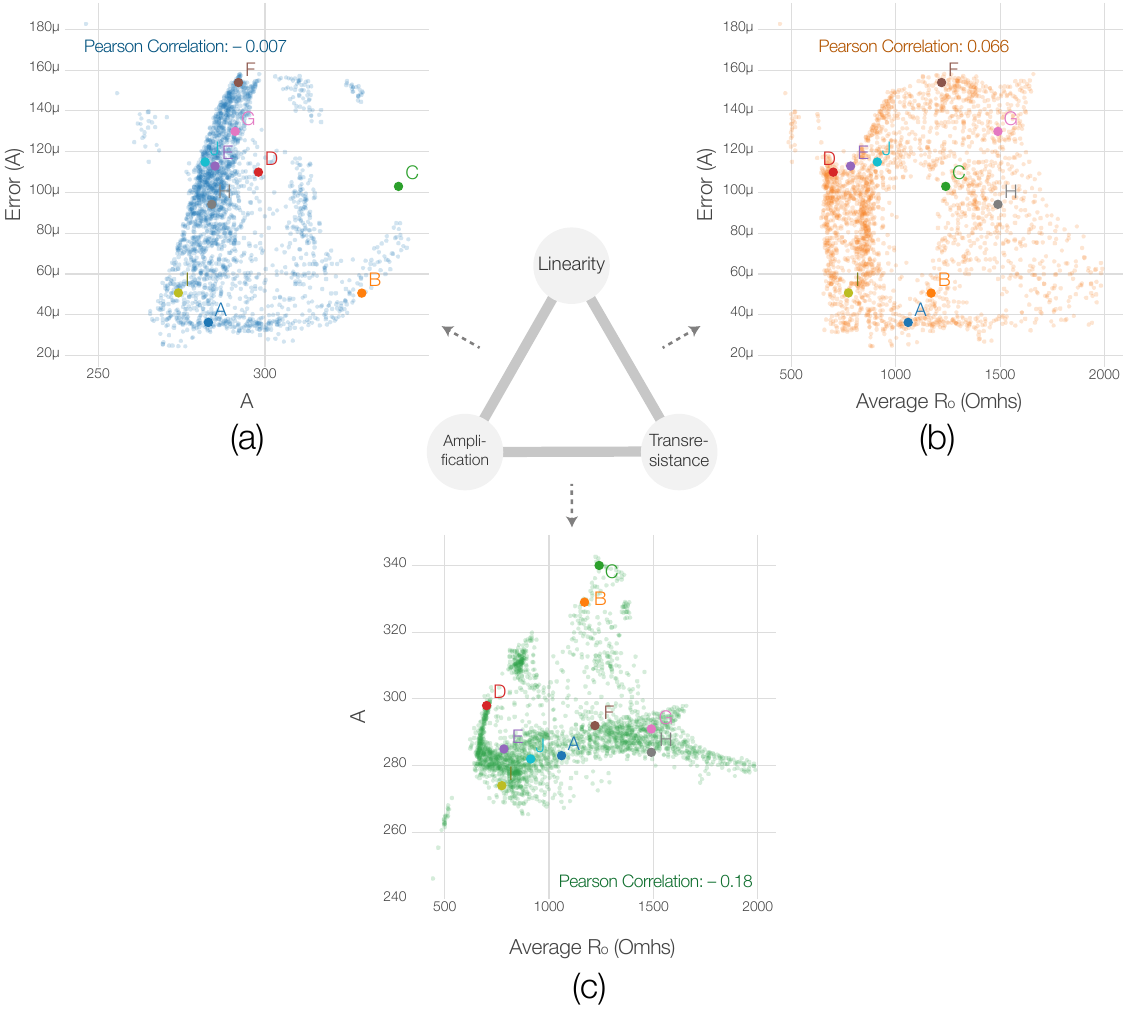} 
  \caption{Relationship between the linearity error and (a) average amplification and (b) average output resistance. The Pearson correlation coefficient between the variables is shown inside each plot. Also shown are the position of the considered load lines.}
  ~\label{f:linScatter}
  \end{center}
\end{figure*}

If a large amplification is desired, the results shown in Figure~\ref{f:RlHTMap}(a) indicate that one should ideally use $V_{cc}$ and $R_c$ close to those of load line B.  For this region, we have near-optimal linearity, large amplification, and moderate output resistance.  Figure~\ref{f:linScatter}(a) shows a scatter-plot between the error and the average amplification, as well as the Pearson correlation coefficient between these two quantities. The obtained Pearson correlation indicates that there is no relationship between these two parameters.   A similar situation is verified for the output resistance, i.e.\ as shown in Figure~\ref{f:linScatter}(b), this constant exhibits weak correlation with the linearity.  Weak correlation was also found between average amplification and output resistance (Figure~\ref{f:linScatter}(c)).  Nevertheless, the three scatterplots in Figure~\ref{f:linScatter} exhibit a surprisingly elaborated structure, suggesting that transistor amplification without feedback is more complex than usually thought.

A traditional way to study the linearity of an amplifier is by estimating its total harmonic distortion (THD)~\cite{cordell2011designing}. For a given frequency $f$, this measurement can be obtained by using a pure sinusoidal function with frequency $f$ as input, identifying new harmonic components in the output (a perfectly linear amplifier would produce no such components), and taking the ratio between the magnitudes of these spurious harmonics ($V_{2f}$, $V_{3f}$, etc) and of the fundamental ($V_f$). More formally, the THD can be calculated as:

\begin{equation}
THD(f) = {\sqrt{V_{2f}^2 + V_{3f}^2 + V_{4f}^2 + \cdots} \over V_f}
\end{equation}

Because the load is purely resistive, the same THD will be attained irrespectively of the input frequency $f$. Therefore, we considered a sinusoidal function with $f=1kHz$. The THD was applied to the chosen load lines indicated in Figure~\ref{f:IaVaMaps}, and the obtained values are shown in Table~\ref{t:parVals}.  A large variation of THD values was found for the chosen load lines, which are a consequence of the variation of the differential geometry of the characteristic surface.

\section{Conclusions}
\label{s:conclusions}

Linear operation has been of paramount importance in most theoretical and applied areas, as a consequence of its ability to preserve the properties of signals, avoiding distortions and other unwanted effects.  Yet, relatively few approaches have been proposed in order to objectively quantify the linearity of a given region of operation in a sensor, device or transducer.  In the present work, we developed a methodology capable of, given a transfer function, finding its respective operation interval allowing maximum linearity.  The reported approach is based on least squares regression, but also incorporates the constraint given by the extent of the desired region of operation.  In addition, all possible intervals are considered, by scanning a window along the domain of the transfer function.  

The methodology has been characterized with respect to the analytical situation involving sigmoid transfer functions in presence of varying levels of noise and also for real-world data related to the properties of a low power, one-stage class A transistor amplifier operating with resistive load.  In the former case, we verified that the method was capable of identifying the optimal region, centered at the origin of the coordinate axis of the sigmoid function, where curvature is known to be smallest.  The application to the amplifier incorporates several contributions, such as the determination of the surface of the transistor operation (i.e. $S(V_c, I_c, I_b)$) by using interpolation, which allowed the detailed estimation of the transistor constants along a domain in the $V_c \times I_c$ space by using partial derivatives, and the estimation of the linearity error in terms of amplification and output resistance.  A surprisingly complex structure was found to underlie the characteristic surface of the adopted small-signal transistor.  Relationships between these three features were then investigated in terms of Pearson correlation coefficients, and it was found that the three properties present low correlation between them.   Such results ultimately allowed the identification of load lines and respective operation points providing a good compromise between high linearity, amplification and low output resistance.

The reported methodology and results provide several possibilities for future investigations.   For instance, it would be interesting to apply the method to optimize the operation of sensors and transducers, as well as of amplifiers involving other configurations and devices (e.g. class AB, vacuum tubes and integrated circuits).  Other linearity criteria could be used, for instance THD.  The surprisingly complex structure of the characteristic surface obtained for the small signal transistor also motivates further investigation, including other models of transistors.  It would also be interesting to develop intelligent control systems using the proposed linearity optimization approach in order to dynamic and interactively set up the best operation points in such devices and systems.  In addition, it should be also observed that, though presented here in the context of electronics and instrumentation, the proposed methodology can be directly used to tackle many important problems in other areas, such as identifying linear regions underlying power-law relationships in logarithmically related measurements (e.g. scale free complex networks~\cite{barabasi1999emergence}).

\section*{Acknowledgements}

F. N. Silva acknowledges FAPESP (Grant No. 15/08003-4). C. H. Comin thanks FAPESP (Grant No. 15/18942-8) for financial support. L. da F. Costa thanks CNPq (Grant no. 307333/2013-2) and NAP-PRP-USP for support. This work has been supported also by FAPESP grant 11/50761-2.

\bibliographystyle{apsrev}
\bibliography{paper}

\section*{Appendix A}

The total current gain of the circuit used for the experiments (shown in Figure~\ref{f:tubeCircuit}) is given by

\begin{equation}
A = \frac{dI_c}{dI_b}.
\end{equation}
Since $I_c$, the collector current, is a function of $V_c$, the collector voltage, and $I_b$, the base current, we can also write $A$ in terms of the partial derivatives of $I_c$, that is

\begin{equation}
A = \frac{dI_c}{dI_b}= \left( \frac{\partial I_c}{\partial V_c}\frac{dV_c}{dI_b} + \frac{\partial I_c}{\partial I_b}\frac{dI_b}{dI_b} \right) .
\end{equation}
Replacing the partial derivatives by the transistor properties indicated in Equations~\ref{eq:beta} and~\ref{eq:ro} we obtain

\begin{equation}
A = \frac{dI_c}{dI_b} = \frac{1}{R_o}\frac{dV_c}{dI_b} + \beta.
\end{equation}
Since $V_c$ and $I_c$ are related through the circuit parameters according to

\begin{equation}
V_{cc} = R_c I_c + V_c ,
\end{equation}
the total derivative $dV_c/dI_b$ can be rewritten in terms of $A$ and $R_c$ as

\begin{equation}
\frac{dV_c}{dI_b} = -R_c\frac{dI_c}{dI_b} = -R_c A.
\end{equation}
Therefore,

\begin{align}
A &= -\frac{R_c}{R_o}A + \beta\\
A &= \frac{R_o\beta}{R_o+R_c}.
\end{align}

%
%
%


\end{document}